\documentclass[11pt]{article}
\usepackage{subcaption, csquotes}

\usepackage[
	style=phys,
	biblabel=brackets,
	eprint=true,
	bibencoding=auto,
	backend=biber,
	sorting=none,
	maxbibnames=3,
	language=auto,
	autolang=other,
	doi=false,
	url=false
	]{biblatex}

\AtEveryBibitem{
	\clearfield{day}
	\clearfield{month}
	\clearfield{endday}
	\clearfield{endmonth}
}

\bibliography{bib.bib}

\usepackage{graphicx}
\usepackage{amsmath}
\usepackage{amssymb}
\usepackage{amsxtra}

\usepackage[colorlinks]{hyperref}

\title{Dynamical evolution of a cluster of primordial black holes}
\author{
    V. D. Stasenko \\
    StasenkoVD@gmail.com \\
    A. A. Kirillov \\
    AAKirillov@mephi.ru \\
    National Research Nuclear University MEPhI \\
    (Moscow Engineering Physics Institute)
    }

\date{}

\begin{document}
\maketitle

\begin{abstract}
Evolution of a cluster of primordial black holes in the two-body relaxation approximation based on the Fokker-Planck equation is discussed. In our calculation, we consider the self-gravitating cluster with a wide range of black holes masses from $10^{-4} M_{\odot}$ up to $100 M_{\odot}$ and the total mass $10^5 M_{\odot}$. Moreover, we included a massive black hole in the cluster center which determines the evolution rate of the density profile in its vicinity.
\end{abstract}



\section{Introduction}\label{s:intro}
The hypothesis of primordial black holes (PBHs) formation was suggested in \cite{1967SvA....10..602Z}. 
Afterward, a few scenarios of PBHs production have been developed (see reviews \cite{2010RAA....10..495K, 2020ARNPS..7050520C}).
In our work, we consider those predicting the formation of PBHs as clusters. This mechanism
was proposed in \cite{2000hep.ph....5271R, 2001JETP...92..921R, 2005APh....23..265K} where a collapse of large closed domain walls was discussed.
The produced clusters may have extended mass spectra where masses range from $\sim10^{17}$~g \cite{2011GrCo...17...27B, 2011APh....35...28B} up to $\sim10^4 M_\odot$ \cite{2008ARep...52..779D} or even more \cite{2005GrCo...11...99D}. These clusters have essential consequences for shedding light on some cosmological problems. Observational manifestations of the model and smoothing of some constraints (the recent restrictions on PBHs are considered in \cite{2020ARNPS..7050520C}) are widely discussed in reviews \cite{2014MPLA...2940005B, 2019EPJC...79..246B} and references within.
However, finding of clusters evidences is significantly related to the mass spectrum at a specified moment of the Universe history. Therefore, understanding of cluster dynamic play an essential role and is a main research subject of this paper.


Till now, a comprehensive study of clusters evolution has not been carried out. First efforts to retrace changes of clusters mass spectra were made in \cite{2019JPhCS1390a2090K, 2019EPJC...79..246B, 2020arXiv200615018T} where N-body simulations of cluster dynamics was discussed. 
The closest physical model to a PBH cluster is a globular cluster of stars. However, it does not have a wide mass range. Therefore, globular cluster theory could not be directly extrapolated to the PBHs cluster case. Moreover, the PBHs cluster may contain a massive central black hole (CBH). In work \cite{1976ApJ...209..214B}, stationary distribution of stars around a massive black hole was discussed. It was established that the density obeys the law $\rho \propto r^{-7/4}$. However, there are a few disadvantages of this work. First, the potential from star distribution is neglected in comparison with the potential of a CBH. Second, all stars have the same mass. Besides, as was shown in \cite{2017ApJ...848...10V, 1982ApJ...253..921D, 2018PhRvD..98b3021S}, a distribution around massive black holes isn't stationary.

We focus on describing the evolution of a PBHs cluster using the orbit-averaged Fokker-Planck equation in energy space. The considered clusters have a wide mass ranges. In addition, we include in our calculations a massive BH in a cluster center which determines behaviour of density profile in a central region. 

\section{The Fokker-Planck equation}

We study the spherically symmetric system of gravitating point masses and assume relaxation time is much longer than an orbital period, and there is no isotropy in velocity space. According to the assumptions, the distribution function (DF) $f$ describing PBHs in a cluster depends only on energy $E$: $f(\textbf{\textit{r}},\textbf{\textit{v}}) = f(E)$. Evolution of the DF is described by the orbit-averaged Fokker-Planck equation which in the multi-mass case has the form \cite{1980ApJ...242..765C, 2017ApJ...848...10V}:
\begin{align}\label{fp_E}
    \frac{\partial N_i}{\partial t} = \frac{\partial}{\partial E} \left (m_i D_{E}(E,f)\, f_i + D_{EE}(E,f)\, \frac{\partial f_i}{\partial E} \right),
\end{align}
where $m_i$, $f_i$ is mass and distribution function of $i$-th type of PBHs, respectively, and $N_i(E,t) = 4 \pi^2 p(E) f_i(E,t)$ is number density in energy space. Expressions for the coefficients in \eqref{fp_E} are
\begin{align}
    D_E(E,f) &= 16 \pi^3 \Gamma \sum_i \int_{\phi(0)}^E m_i f_i(E') p(E') \, dE',
    \\
    D_{EE}(E,f) &= 16 \pi^3 \Gamma \sum_i \Biggl(  q(E) \int_{E}^0 m^2_i f_i(E') \, dE ' +
    \notag
    \\
    &+ \int_{\phi(0)}^{E} m^2_i f_i(E') q(E') \ dE' \Biggr),
    \label{DEE}
\end{align}
where the sum goes over all types of masses and $\Gamma = 4 \pi G^2 \ln{\Lambda}$, $\ln{\Lambda}$ is the Coulomb logarithm.
$p(E)$ and $q(E)$ are given by
\begin{equation}
    \begin{aligned}
        \label{qp}
        p(E) & = 4 \int_0^{\phi^{-1}(E)} dr \, r^2 \sqrt{2 (E - \phi(r))},
        \\
        q(E) & = \frac{4}{3} \int_{0}^{\phi^{-1}(E)} dr \, r^2  \Big[2(E - \phi(r)) \Big]^{3/2},
    \end{aligned}
\end{equation}
where $\phi^{-1}(E)$ is the root of the equation $E = \phi(r)$. Asymptotic expressions for $q(E)$ and $p(E)$ in the case of the Keplerian potential $\phi(r) = -GM_{\bullet}/r$ is:
\begin{align} \label{qE_pE_asymp}
    p(E) = \frac{\sqrt{2} \pi (G M_{\bullet})^3}{4 (-E)^{5/2}} = -\frac{3 q}{2 E},
    \quad
    q(E) = \frac{\sqrt{2} \pi (G M_{\bullet})^3}{6 (-E)^{3/2}},
\end{align}
which can be used to calculate $p(E)$ and $q(E)$ near the CBH or at large distances from the cluster. Here, $M_{\bullet}$ is the CBH mass.

To study evolution of a self-gravitating system, it is necessary to solve together the Fokker-Planck equation \eqref{fp_E} and the Poisson equation
\begin{align} \label{poisson}
    \phi(r) = - 4 \pi G \left ( \frac{1}{r} \int_0^r dr' \, r'^2 \rho(r') + \int_r^{\infty} dr' \, r' \rho(r') \right) - \frac{G M_{\bullet}}{r},
\end{align}
where $\rho(r)$ is given by the expression
\begin{align} \label{rho}
    \rho(r) = 4 \pi \sum_{i} m_i \int_{\phi(r)}^0 dE \, f_i(E) \sqrt{2(E - \phi(r))} = \sum_{i} \rho_i(r).
\end{align}
The technique of joint solution of the Poisson equation \eqref{poisson} and the Fokker-Planck equation \eqref{fp_E} was described in the work \cite{1980ApJ...242..765C} and improved in \cite{2017ApJ...848...10V}.

In practice, the density profile is initially defined, not the distribution function. In order to obtain the initial distribution function, it is necessary to use the Eddington formula \cite{2008gady.book.....B,2013degn.book.....M}:
\begin{align} \label{Eddington}
    f_{i}(E) = \frac{\sqrt{2}}{4 \pi^2 m_i} \frac{d}{d E} \int_{E}^{0} \frac{d \phi}{\sqrt{\phi - E}} \frac{d \rho_{i}}{d \phi}.
\end{align}

\section{Results}
We choose the initial density profile of the cluster in the form:
\begin{align}
    \rho_i(r) = \rho_{0,i} \left ( \frac{r}{r_0} \right)^{-2} \left [ 1 + \left ( \frac{r}{r_0}\right)^{2} \right]^{-3/2},
\end{align}
where $\rho_{0,i}$ is the normalization factor and $r_{0} = 0.5$~pc. The mass spectrum is the same as in \cite{2019EPJC...79..246B}
\begin{align}
    \frac{dN}{dM} \propto \frac{1}{M_{\odot}} \left ( \frac{M}{M_{\odot}} \right)^{-2},
\end{align}
and PBHs masses range from $10^{-4} M_{\odot}$ up to $10^2 M_{\odot}$. We take the bin width such that the total masses of each component of PBHs are equal to each other. On a logarithmic scale, this corresponds to the same bin widths for the spectrum $d N / dM \propto M^{-2}$. Number of PBHs types is 10. The total mass of the cluster is $10^{5} M_{\odot}$, i.e. the total mass of each component is $10^{4} M_{\odot}$. The mass of the CBH is $ M_{\bullet} = 10^{3} M_{\odot}$

After we define the initial density profile and mass spectrum, we could solve the equations \eqref{fp_E} and \eqref{poisson}. The solution technique can be found in \cite{1980ApJ...242..765C, 2017ApJ...848...10V}. Here we present the results.

\begin{figure}[!t]
	\centering
	\begin{subfigure}[t]{0.48\textwidth}
		\includegraphics[width = \textwidth]{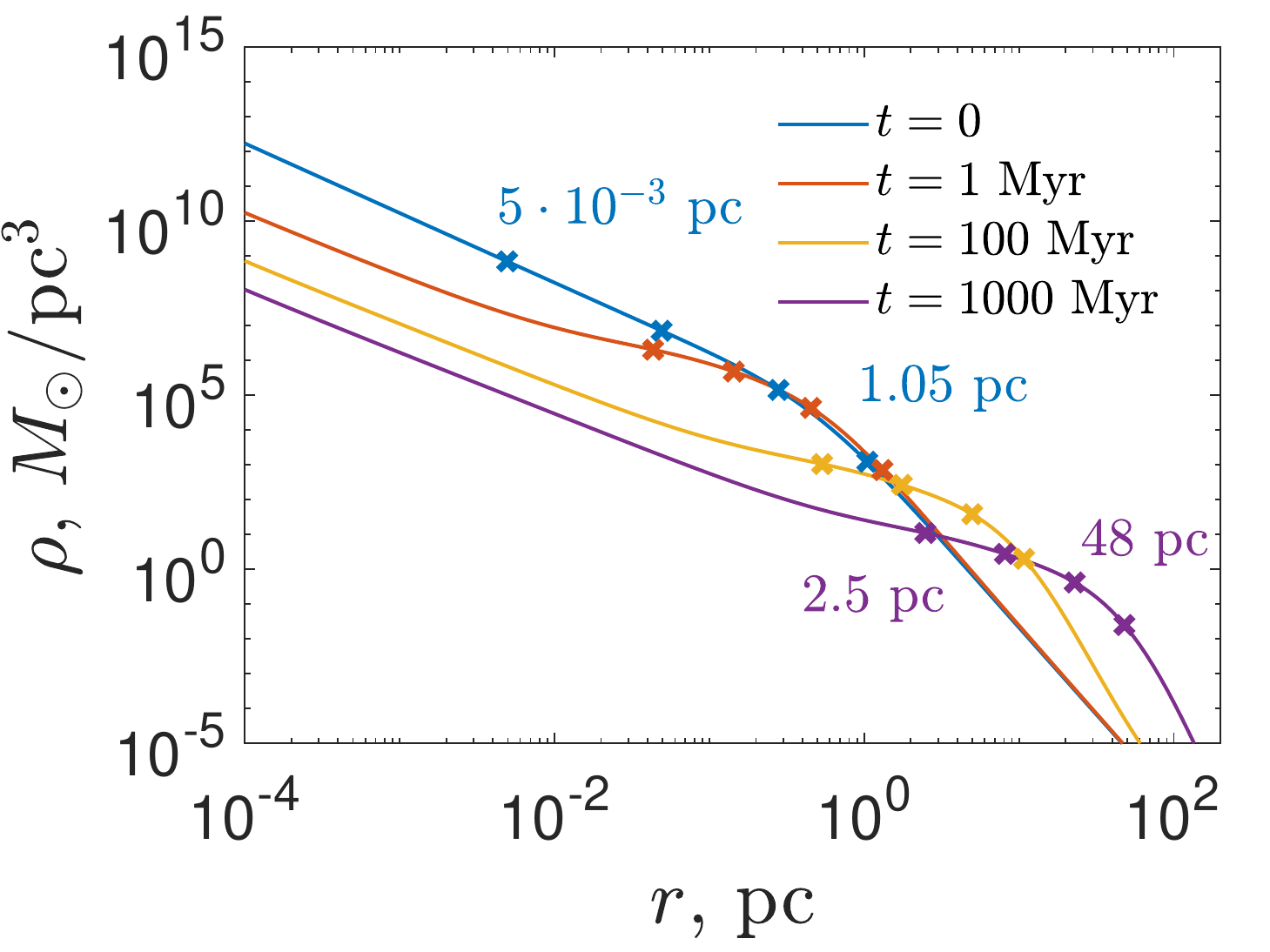}
		\caption{The PBHs density profile in the cluster over time.}
		\label{rho_tot}
	\end{subfigure}
	\hfil
	\begin{subfigure}[t]{0.48\textwidth}
		\includegraphics[width = \textwidth]{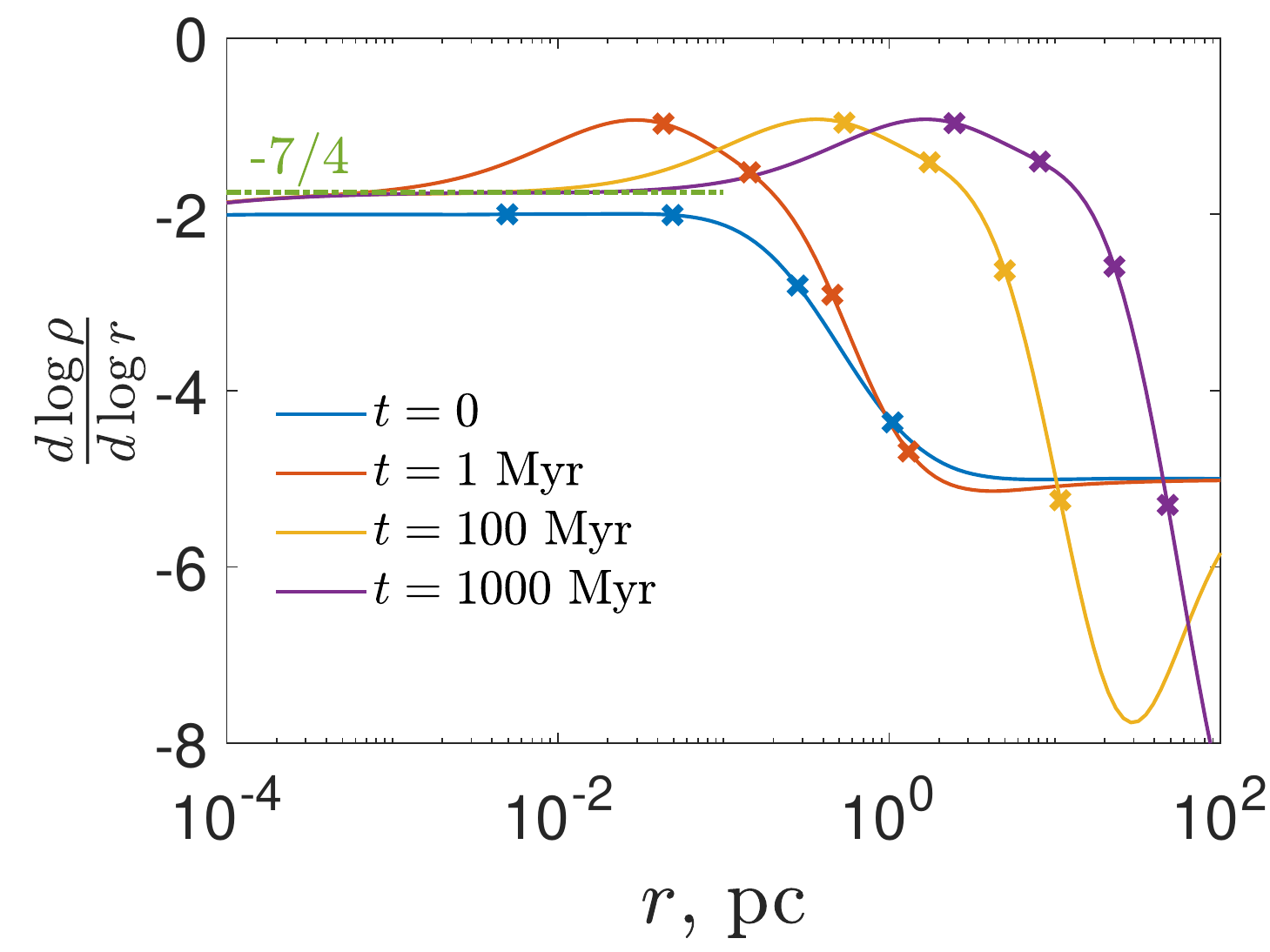}
		\caption{The slope of the density profile.}
		\label{slope_tot}
	\end{subfigure}
	\begin{subfigure}[t]{0.48\textwidth}
		\includegraphics[width = \textwidth]{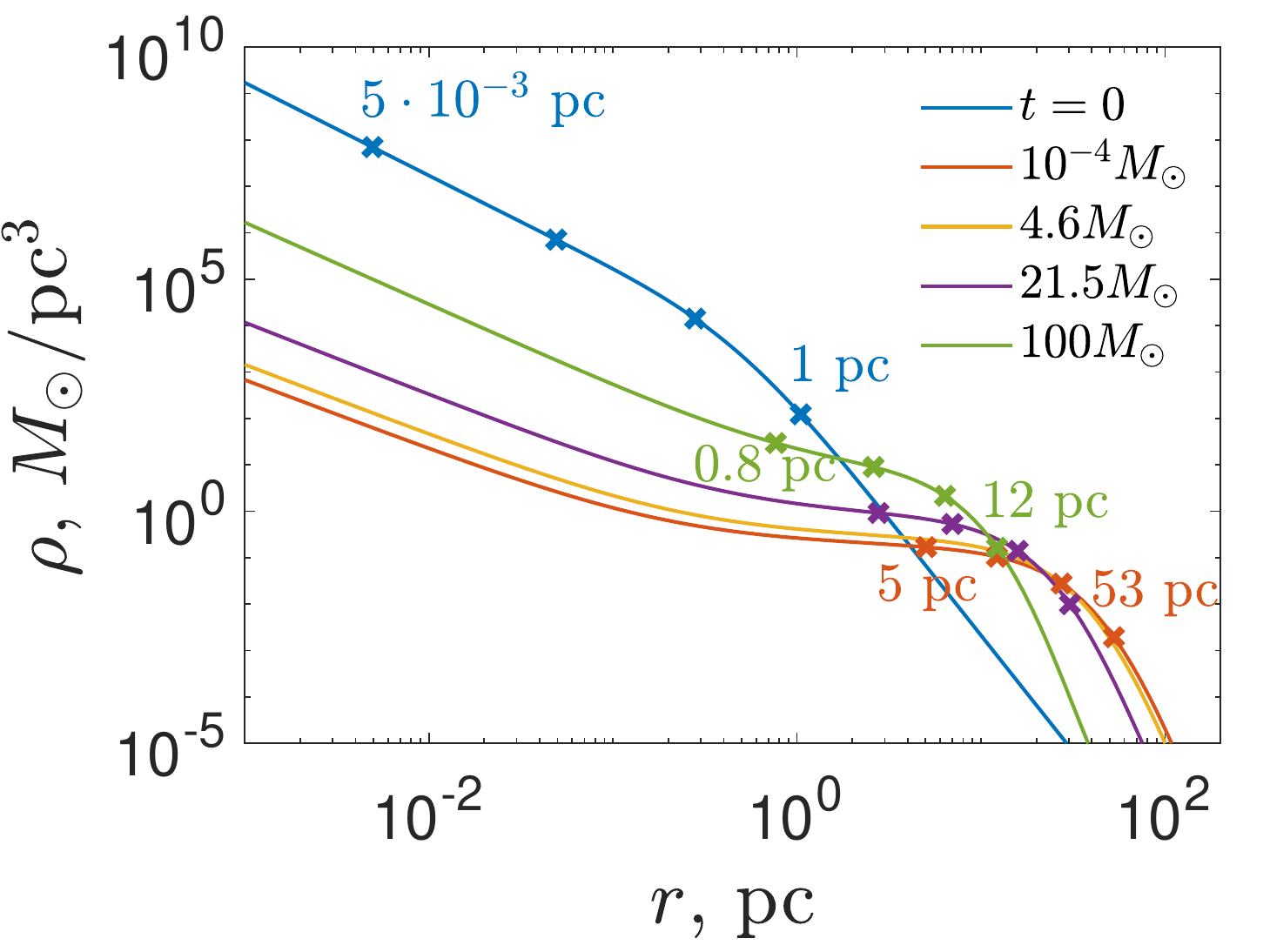}
		\caption{The density profiles of the different types of PBHs at the initial and the final times.}
		\label{rho_comp}
	\end{subfigure}
	\hfil
	\begin{subfigure}[t]{0.48\textwidth}
		\includegraphics[width = \textwidth]{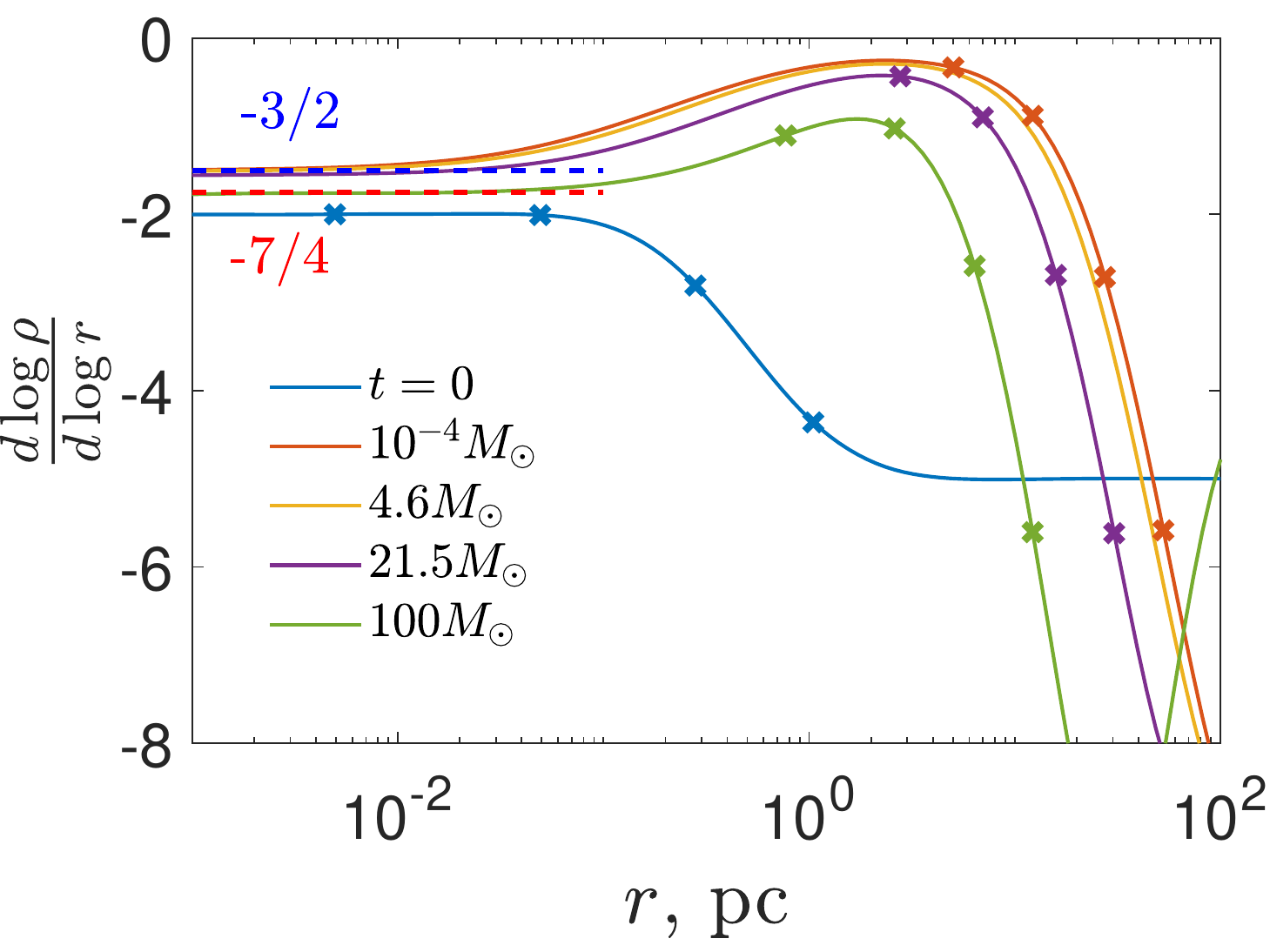}
		\caption{The slopes of the density profiles for the different types of PBHs at the initial and the final times.}
		\label{slope_comp}
	\end{subfigure}
	\caption{The evolution of the mass density of PBHs in the cluster is shown. Crosses corresponds to radii containing 1, 10, 50 90 \% (from left to right) of the total mass. The radius values refer to the first and the last crosses at the initial and the final ($\sim1$~Gyr) times.}
	\label{Rho_tot_evolv}
\end{figure}

Figure~\ref{Rho_tot_evolv} shows the evolution of the total density profile of PBHs and its slope. After $\sim$1 Myr, the cusp $\rho \propto r^{-7/4}$ is established in the central region, it is well known as the Bachall-Wolf cusp \cite{1976ApJ...209..214B}. Over time, the behaviour of the density profile extends to a larger values of radii. However, in figure~\ref{slope_tot}, one can see that only a small part of the cluster ($< 1 \%$) has the profile $\rho \propto r^{-7/4}$. It can also be seen the cluster size has grown by $\sim 50$ times.

In figure~\ref{rho_comp}, it is shown that the behavior of the density profile in the central region of the cluster is determined by the most massive PBHs.
Moreover, one can see heavy PBHs tend to be located near the central region of the cluster while the light ones tend to be at the periphery. Figure~\ref{slope_comp} presents that by the final time $\sim1$~Gyr, the density profile of the heaviest component obtains the cusp in the central region $\rho \propto r^{-7/4}$ and the lighter components achieve $\rho \propto r^{-3/2}$.

\begin{figure}[!t]
	\centering
	\begin{subfigure}[t]{0.48\textwidth}
		\includegraphics[width = \textwidth]{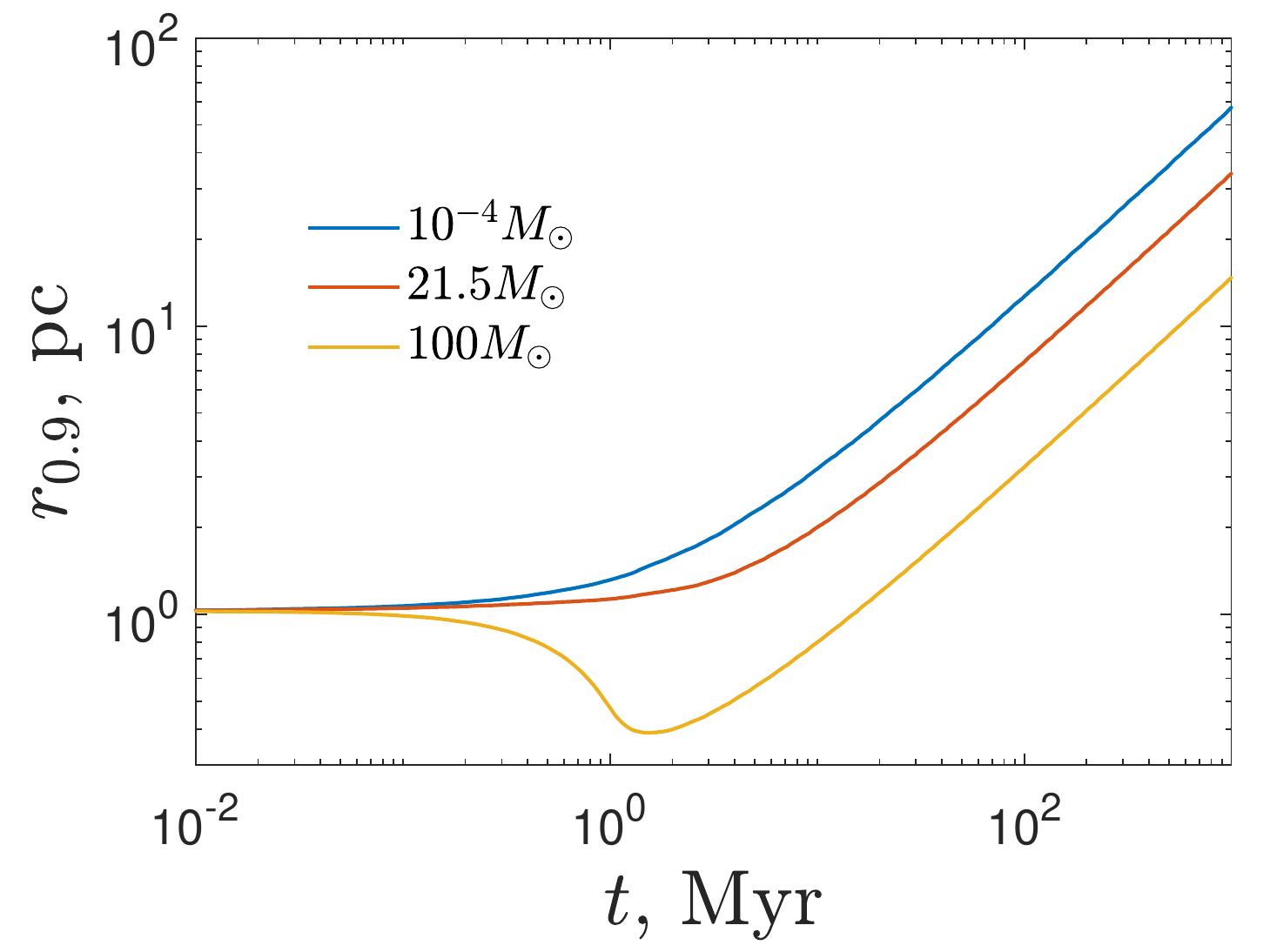}
		\caption{The time dependence of the radii of the spheres containing 90\% of the total mass for the different types of PBHs.}
		\label{comp_radii}
	\end{subfigure}
	\hfil
	\begin{subfigure}[t]{0.48\textwidth}
		\includegraphics[width = \textwidth]{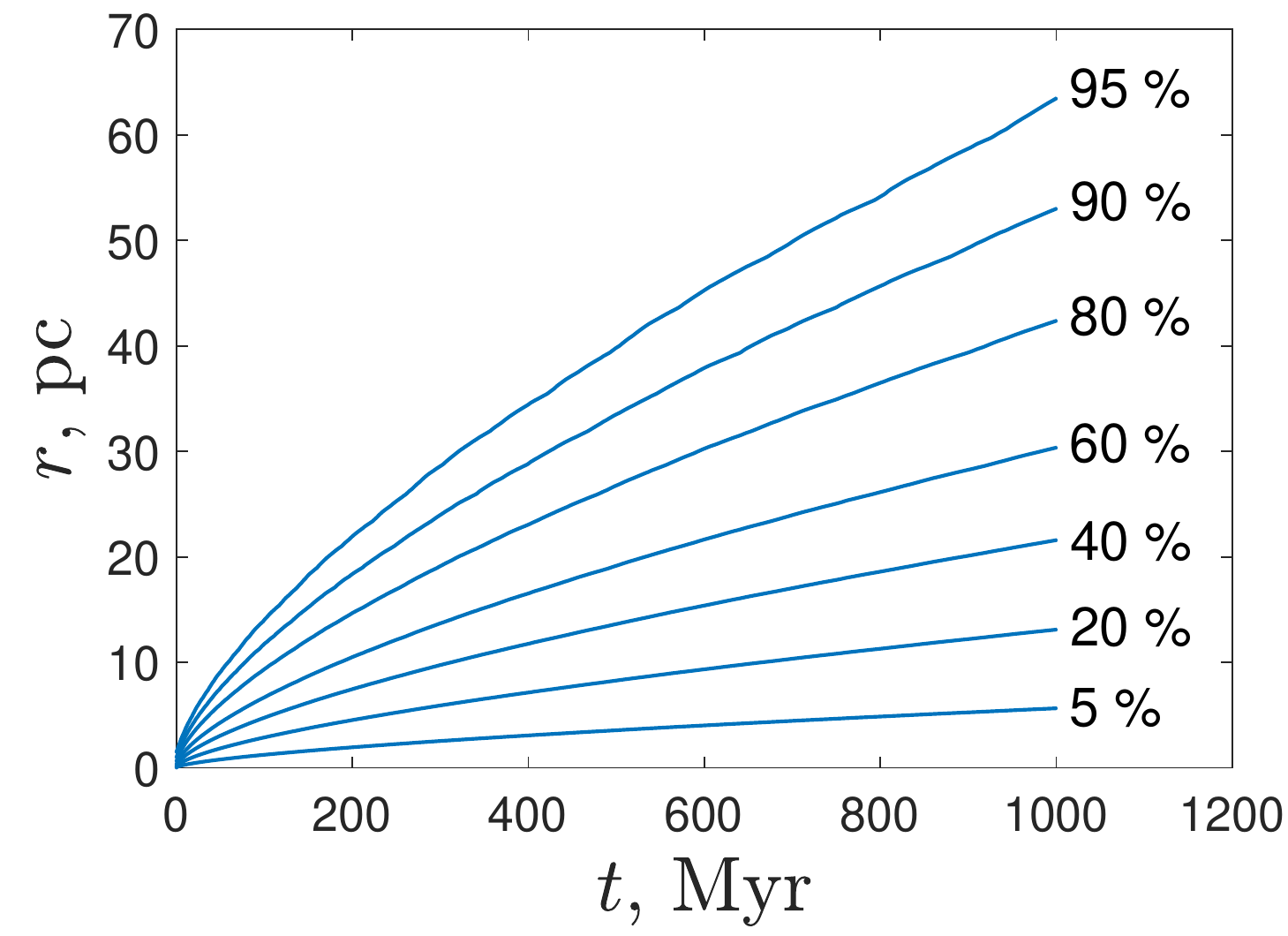}
		\caption{The evolution of the radii of the sphere containing the indicated percentage of the total cluster mass.}
		\label{total_radii}
	\end{subfigure}
	\caption{The evolution of the mass distribution.}
	\label{Radii}
\end{figure}

In order to show how the redistribution of masses occurs, figure~\ref{comp_radii} illustrates the time dependence of the radii $r_{0.9}$ of the spheres containing 90\% of masses of each component. It can be see that at first $~1$~Myr, heavy PBHs are compressed toward the center of the cluster due to dynamical friction while other ones evolve in ``a slow mode''. Then, the cluster begins to expand. Figure~\ref{total_radii} shows the time dependence of the sphere radius for the whole cluster. Thus, one can find a significant increase in the cluster size by $\sim 50$ times over $\sim1$~Gyr.

\section*{Conclusion}
In this paper, the evolution of the PBHs cluster within the Fokker-Plank framework is studied. We note the significant redistribution of the cluster structure leading heavy black holes tend to be located near the cluster center and lighter ones tend to be at the periphery. In addition, a significant increase in the cluster size by $\sim 50$ times for a given initial density profile and mass spectrum is shown.

At the end of the conclusion, it should be noted the growth of the CBH mass due to gas accretion and capture of surrounding PBHs may change the cluster evolution rate, but it is beyond the scope of this research and should be considered separately.

\section*{Acknowledgements}
The authors are grateful to K.~M.~Belotsky and S.~G.~Rubin for useful discussions. The work was supported by the Ministry of Science and Higher Education of the Russian Federation, project ``Fundamental properties of elementary particles and cosmology'' No 0723-2020-0041.

\printbibliography

@string{annurev="Annu. Rev. Nucl. Part. Sci. "}

@string{apj="Astrophys. J. "}

@string{arep="Astron. Rep. "}

@string{astropart="Astropart. Phys. "}

@string{epjc ="Eur. Phys. J. C"}

@string{gc="Grav. Cosmol. "}

@string{jetp="J. Exp. Theor. Phys. "}

@string{jpcs="J. Phys. Conf. Ser. "}

@string{mpla="Mod. Phys. Lett. A "}

@string{prd="Phys. Rev.~D "}

@string{raa="Res. Astron. Astrophys. "}

@string{sovast="Sov. Astron. "}

@ARTICLE{1967SvA....10..602Z,
       author = {{Zel'dovich}, Ya. B. and {Novikov}, I.~D.},
        title = "{The Hypothesis of Cores Retarded during Expansion and the Hot Cosmological Model}",
      journal = sovast,
         year = 1967,
        month = feb,
       volume = {10},
        pages = {602},
       adsurl = {https://ui.adsabs.harvard.edu/abs/1967SvA....10..602Z}
}

@ARTICLE{2001JETP...92..921R,
       author = {{Rubin}, S.~G. and {Sakharov}, A.~S. and {Khlopov}, M. Yu.},
        title = "{The Formation of Primary Galactic Nuclei during Phase Transitions in the Early Universe}",
      journal = jetp,
         year = 2001,
        month = jun,
       volume = {92},
       number = {6},
        pages = {921-929},
          doi = {10.1134/1.1385631},
archivePrefix = {arXiv},
       eprint = {hep-ph/0106187},
 primaryClass = {hep-ph},
       adsurl = {https://ui.adsabs.harvard.edu/abs/2001JETP...92..921R}
}

@ARTICLE{2005APh....23..265K,
       author = {{Khlopov}, M. Yu. and {Rubin}, S.~G. and {Sakharov}, A.~S.},
        title = "{Primordial structure of massive black hole clusters}",
      journal = astropart,
         year = 2005,
        month = mar,
       volume = {23},
       number = {2},
        pages = {265-277},
          doi = {10.1016/j.astropartphys.2004.12.002},
archivePrefix = {arXiv},
       eprint = {astro-ph/0401532},
 primaryClass = {astro-ph},
       adsurl = {https://ui.adsabs.harvard.edu/abs/2005APh....23..265K}
}

@ARTICLE{1976ApJ...209..214B,
       author = {{Bahcall}, J.~N. and {Wolf}, R.~A.},
        title = "{Star distribution around a massive black hole in a globular cluster.}",
      journal = apj,
         year = 1976,
        month = oct,
       volume = {209},
        pages = {214-232},
          doi = {10.1086/154711},
       adsurl = {https://ui.adsabs.harvard.edu/abs/1976ApJ...209..214B}
}

@ARTICLE{2017ApJ...848...10V,
       author = {{Vasiliev}, Eugene},
        title = "{A New Fokker-Planck Approach for the Relaxation-driven Evolution of Galactic Nuclei}",
      journal = apj,
         year = 2017,
        month = oct,
       volume = {848},
       number = {1},
        pages = {10},
          doi = {10.3847/1538-4357/aa8cc8},
archivePrefix = {arXiv},
       eprint = {1709.04467},
 primaryClass = {astro-ph.GA},
       adsurl = {https://ui.adsabs.harvard.edu/abs/2017ApJ...848...10V}
}

@BOOK{2008gady.book.....B,
       author = {{Binney}, James and {Tremaine}, Scott},
        title = "{Galactic Dynamics: Second Edition}",
         year = 2008,
         publisher = {Princeton University Press},
         address = {Princeton},
       adsurl = {https://ui.adsabs.harvard.edu/abs/2008gady.book.....B}
}

@ARTICLE{1980ApJ...242..765C,
       author = {{Cohn}, H.},
        title = "{Late core collapse in star clusters and the gravothermal instability}",
      journal = apj,
         year = 1980,
        month = dec,
       volume = {242},
        pages = {765-771},
          doi = {10.1086/158511},
       adsurl = {https://ui.adsabs.harvard.edu/abs/1980ApJ...242..765C}
}

@BOOK{2013degn.book.....M,
       author = {{Merritt}, David},
        title = "{Dynamics and Evolution of Galactic Nuclei}",
         year = 2013,
         publisher = {Princeton University Press},
         address = {Princeton},
       adsurl = {https://ui.adsabs.harvard.edu/abs/2013degn.book.....M}
}

@ARTICLE{2000hep.ph....5271R,
       author = {{Rubin}, S.~G. and {Khlopov}, M. Yu. and {Sakharov}, A.~S.},
        title = "{Primordial Black Holes from Non-Equilibrium Second Order Phase Transition}",
      journal = gc,
         year = 2000,
        month = may,
       volume = {S6},
        pages = {51-58},
archivePrefix = {arXiv},
       eprint = {hep-ph/0005271},
 primaryClass = {hep-ph},
       adsurl = {https://ui.adsabs.harvard.edu/abs/2000hep.ph....5271R}
}

@ARTICLE{2005GrCo...11...99D,
       author = {{Dokuchaev}, V.~I. and {Eroshenko}, Yu. N. and {Rubin}, S.~G.},
        title = "{Quasars formation around clusters of primordial black holes}",
      journal = gc,
         year = 2005,
        month = jun,
       volume = {11},
        pages = {99-104},
archivePrefix = {arXiv},
       eprint = {astro-ph/0412418},
 primaryClass = {astro-ph},
       adsurl = {https://ui.adsabs.harvard.edu/abs/2005GrCo...11...99D}
}

@ARTICLE{2008ARep...52..779D,
       author = {{Dokuchaev}, V.~I. and {Eroshenko}, Yu. N. and {Rubin}, S.~G.},
        title = "{Early formation of galaxies induced by clusters of black holes}",
      journal = arep,
         year = 2008,
        month = oct,
       volume = {52},
       number = {10},
        pages = {779-789},
          doi = {10.1134/S1063772908100016},
archivePrefix = {arXiv},
       eprint = {0801.0885},
 primaryClass = {astro-ph},
       adsurl = {https://ui.adsabs.harvard.edu/abs/2008ARep...52..779D}
}

@ARTICLE{2010RAA....10..495K,
       author = {{Khlopov}, Maxim Yu.},
        title = "{Primordial black holes}",
      journal = raa,
         year = 2010,
        month = jun,
       volume = {10},
       number = {6},
        pages = {495-528},
          doi = {10.1088/1674-4527/10/6/001},
archivePrefix = {arXiv},
       eprint = {0801.0116},
 primaryClass = {astro-ph},
       adsurl = {https://ui.adsabs.harvard.edu/abs/2010RAA....10..495K}
}

@ARTICLE{2011GrCo...17...27B,
       author = {{Belotsky}, K.~M. and {Berkov}, A.~V. and {Kirillov}, A.~A. and
         {Rubin}, S.~G.},
        title = "{Black hole clusters in our Galaxy}",
      journal = gc,
         year = 2011,
        month = jan,
       volume = {17},
       number = {1},
        pages = {27-30},
          doi = {10.1134/S020228931101004X},
       adsurl = {https://ui.adsabs.harvard.edu/abs/2011GrCo...17...27B}
}

@ARTICLE{2011APh....35...28B,
       author = {{Belotsky}, K.~M. and {Berkov}, A.~V. and {Kirillov}, A.~A. and
         {Rubin}, S.~G.},
        title = "{Clusters of black holes as point-like gamma-ray sources}",
      journal = astropart,
         year = 2011,
        month = aug,
       volume = {35},
       number = {1},
        pages = {28-32},
          doi = {10.1016/j.astropartphys.2011.05.001},
archivePrefix = {arXiv},
       eprint = {1212.2524},
 primaryClass = {astro-ph.HE},
       adsurl = {https://ui.adsabs.harvard.edu/abs/2011APh....35...28B}
}

@ARTICLE{2014MPLA...2940005B,
       author = {{Belotsky}, K.~M. and {Dmitriev}, A.~E. and {Esipova}, E.~A. and
         {Gani}, V.~A. and {Grobov}, A.~V. and {Khlopov}, M. Yu. and
         {Kirillov}, A.~A. and {Rubin}, S.~G. and {Svadkovsky}, I.~V.},
        title = "{Signatures of primordial black hole dark matter}",
      journal = mpla,
         year = 2014,
        month = nov,
       volume = {29},
       number = {37},
          eid = {1440005},
          doi = {10.1142/S0217732314400057},
archivePrefix = {arXiv},
       eprint = {1410.0203},
 primaryClass = {astro-ph.CO},
       adsurl = {https://ui.adsabs.harvard.edu/abs/2014MPLA...2940005B}
}

@ARTICLE{2019EPJC...79..246B,
       author = {{Belotsky}, Konstantin M. and {Dokuchaev}, Vyacheslav I. and
         {Eroshenko}, Yury N. and {Esipova}, Ekaterina A. and
         {Khlopov}, Maxim Yu. and {Khromykh}, Leonid A. and {Kirillov}, Alexand
        er A. and {Nikulin}, Valeriy V. and {Rubin}, Sergey G. and
         {Svadkovsky}, Igor V.},
        title = "{Clusters of Primordial Black Holes}",
      journal = epjc,
         year = 2019,
        month = mar,
       volume = {79},
       number = {3},
        pages = {246},
          doi = {10.1140/epjc/s10052-019-6741-4},
archivePrefix = {arXiv},
       eprint = {1807.06590},
 primaryClass = {astro-ph.CO},
       adsurl = {https://ui.adsabs.harvard.edu/abs/2019EPJC...79..246B}
}

@article{2019JPhCS1390a2090K,
       author = {{Khromykh}, Leonid A. and {Kirillov}, Alexander A.},
        title = "{The gravitational dynamics of the primordial black holes cluster}",
    journal = jpcs,
         year = 2019,
       volume = {1390},
        month = nov,
          eid = {012090},
          doi = {10.1088/1742-6596/1390/1/012090},
       adsurl = {https://ui.adsabs.harvard.edu/abs/2019JPhCS1390a2090K}
}

@ARTICLE{2020ARNPS..7050520C,
       author = {{Carr}, Bernard and {K{\"u}hnel}, Florian},
        title = "{Primordial Black Holes as Dark Matter: Recent Developments}",
      journal = annurev,
         year = 2020,
        month = oct,
       volume = {70},
       number = {1},
        pages = {355-394},
          doi = {10.1146/annurev-nucl-050520-125911},
archivePrefix = {arXiv},
       eprint = {2006.02838},
 primaryClass = {astro-ph.CO},
       adsurl = {https://ui.adsabs.harvard.edu/abs/2020ARNPS..7050520C}
}

@ARTICLE{2020arXiv200615018T,
       author = {{Trashorras}, Manuel and {Garc{\'i}a-Bellido}, Juan and
         {Nesseris}, Savvas},
        title = "{The clustering dynamics of primordial black boles in $ N $-body simulations}",
archivePrefix = {arXiv},
       eprint = {2006.15018},
 primaryClass = {astro-ph.CO},
       adsurl = {https://ui.adsabs.harvard.edu/abs/2020arXiv200615018T}
}

@ARTICLE{2018PhRvD..98b3021S,
       author = {{Shapiro}, Stuart L.},
        title = "{Star clusters, self-interacting dark matter halos, and black hole cusps: The fluid conduction model and its extension to general relativity}",
      journal = prd,
         year = 2018,
        month = jul,
       volume = {98},
       number = {2},
          eid = {023021},
          doi = {10.1103/PhysRevD.98.023021},
archivePrefix = {arXiv},
       eprint = {1809.02618},
 primaryClass = {astro-ph.HE},
       adsurl = {https://ui.adsabs.harvard.edu/abs/2018PhRvD..98b3021S}
}

@ARTICLE{1982ApJ...253..921D,
       author = {{Duncan}, M.~J. and {Shapiro}, S.~L.},
        title = "{Star clusters containing massive, central black holes. IV - Galactic tidal fields}",
      journal = apj,
         year = 1982,
        month = feb,
       volume = {253},
        pages = {921-938},
          doi = {10.1086/159691},
       adsurl = {https://ui.adsabs.harvard.edu/abs/1982ApJ...253..921D}
}

\end{document}